# AccFFT: A library for distributed-memory FFT on CPU and GPU architectures

Amir Gholami[a,*], Judith Hill[b], Dhairya Malhotra[a], George Biros[a]

*[a]Institute for Computational Engineering and Sciences,*
*The University of Texas at Austin, Austin, TX 78712, USA*
*[b]Computer Science and Mathematics Division,*
*Oak Ridge National Laboratory, Oak Ridge, TN 37831, USA*

## Abstract

We present a new library for parallel distributed Fast Fourier Transforms (FFT). The importance of FFT in science and engineering and the advances in high performance computing necessitate further improvements. AccFFT extends existing FFT libraries for CUDA-enabled Graphics Processing Units (GPUs) to distributed memory clusters. We use overlapping communication method to reduce the overhead of PCIe transfers from/to GPU. We present numerical results on the Maverick platform at the Texas Advanced Computing Center (TACC) and on the Titan system at the Oak Ridge National Laboratory (ORNL). We present the scaling of the library up to 4,096 K20 GPUs of Titan.

*Keywords:* Fast Fourier Transform, Parallel FFT, Distributed FFT, slab decomposition, pencil decomposition

## 1. Introduction

Fast Fourier Transform is one of the most fundamental algorithms in computational science and engineering. It is used in turbulence simulations [20], computational chemistry and biology [8], gravitational interactions [3], cardiac electro-physiology [6], cardiac mechanics [22], acoustic, seismic and electromagnetic scattering [5, 30], materials science [23], molecular docking [19] and many other areas.

Due to its wide range of applications and the need for scalability and performance, the design of FFT algorithms remains an active area of research. Highly optimized single-node FFT algorithms have been implemented by all major hardware vendors, including Intel's MKL library [39], IBM's ESSL library [10], NVIDIA's CUFFT [26] library, and the new AMD's clFFT library [1].

A thorough review of the key challenges to get good performance from single node FFT implementations can be found in [11]. In the realm of open-source software, one of the most widely used libraries is the FFTW [14, 13]. Getting good single-node performance from FFT that works optimally for all platforms is challenging. Therefore, libraries such as FFTW or SPIRAL use auto-tuning or search and learn techniques to find an optimal algorithm for a given platform [33]. Single-node implementations of these libraries have been extended to distributed memory versions either by the original developers or by other research groups. A large number of distributed memory libraries is currently available for CPUs.

*Related work.* There is a vast literature on algorithms for FFTs. Our discussion is by no means exhaustive. We limit it on the work that is most closely related to ours. Introductory material on distributed memory FFT can be found in [16]. Excellent discussions on complexity and performance analysis for 3-D FFTs can be found in [15] and [7].

*[*]Corresponding author
*Email addresses:* `gholami@accfft.org` (Amir Gholami), `hilljc@ornl.gov` (Judith Hill), `dhairya.malhotra@gmail.com` (Dhairya Malhotra), `gbiros@acm.org` (George Biros)

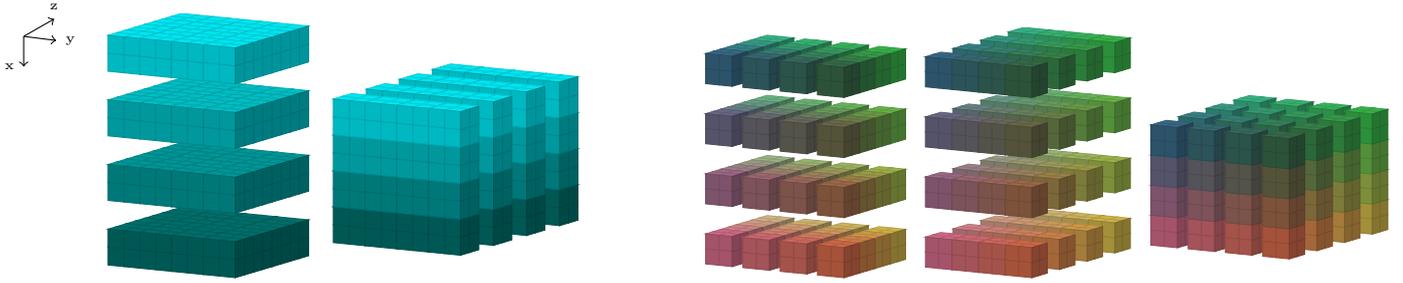

Figure 1: Decomposition of the input/output array in different stages of the forward FFT algorithm. Left: slab decomposition. Right: pencil decomposition

- (*Libraries for CPU architectures.*) One of the most widely used packages for FFTs is the FFTW [14] library. FFTW supports MPI using slab decomposition and hybrid parallelism using OpenMP. However, the scalability of slab decompositions is limited. Furthermore, FFTW does not support GPUs. P3DFFT [29] extends the single-node FFTW (or ESSL) and supports both slab and pencil decompositions. Another recent library is PFFT [31]. The library is built on top of FFTW and uses its transpose functions to perform the communication phase. It has recently been extended to nonequispaced FFTs [32]. It supports distributed multidimensional FFTs, as well as features such as ghost points or pruned transforms. PFFT has an auto-tuning function for finding an optimal communication pattern. A very similar code to P3DFFT and PFFT is 2DECOMP [21] and OpenFFT [9]. These are all very well written libraries that have been used extensively.

A multithreaded code (not open source) is described in [20] in the context of turbulence simulations. This code is based on FFTW and employs single-node optimizations. To our knowledge, this code is one of the most scalable 3-D FFTs. The authors report results on up to 786,432 cores on an IBM Blue Gene machine. However, the authors observe lack of scalability of the transpose for large core counts. On Stampede they start losing scalability at 4,096 nodes. In [35] the authors propose pencil decomposition optimizations that deliver 1.8× speed-up over FFTW. The main idea is the use of non-blocking MPI all-to-all operations that allow overlapping computation and communication. However the method does not address the scalability issues of FFTs. The authors compare FFTW, P3DFFT and 2DECOMP with their scheme. Other works that study 3-D FFTs on x86 platforms include [4, 27].

- (*Libraries for distributed-memory GPUs.*) The work presented in [25] is, to our knowledge, one the most efficient and more scalable distributed GPU implementations. It only supports slab decomposition so it cannot be scaled to large core counts. The scaling results presented in the paper are up to 768 GPUs. The authors employ special techniques to improve the complexity of the transpose and use an in-house CUDA FFT implementation. Their optimizations are specific to the infiniband-interconnect using the IBverbs library and thus is not portable. In the largest run, they observed 4.8TFLOPS for a $2048^3$ problem on 786 M2050 Fermi GPUs (double precision, complex-to-complex), which is roughly 1.2% of the peak performance.

The ohter recent GPU library is digpuFFT [7] which is a modification of P3DFFT in which the intranode computations are replaced by CUFFT. They achieve about 0.7% of the peak. However, digpuFFT code was not designed for production but for experimental validation of the theoretical analysis of the complexity of 3-D FFTs, and its implications to the design of exascale architectures.

- (*1D FFT and single-node libraries.*) Other works that analyze scalability of the FFT codes include the FFTE code [37], which is part of the HPCC benchmark. It includes several optimizations but has support for GPU only via PGI compiler directives. In [40, 17], the authors pro-pose schemes for single node 3-D FFTs. In [24], shared-memory multiple GPU algorithms are discussed. More specialized and somewhat machine-dependent codes are [18] and [12].

**Algorithm 1:** Forward and backward FFT algorithm for pencil decomposition.

| | |
|---|---|
| **Input** : Data in spatial domain.<br>Layout: $N_0/P_0 \times N_1/P_1 \times N_2$ | **Input** : Data in frequency domain.<br>Layout: $\widehat{N}_0 \times \widehat{N}_1/P_0 \times \widehat{N}_2/P_1$ |
| **Output**: Data in frequency domain.<br>Layout: $\widehat{N}_0 \times \widehat{N}_1/P_0 \times \widehat{N}_2/P_1$ | **Output**: Data in spatial domain.<br>Layout: $N_0/P_0 \times N_1/P_1 \times N_2$ |
| $N_0/P_0 \times N_1/P_1 \times \widehat{N}_2 \xleftarrow{FFT} N_0/P_0 \times N_1/P_1 \times N_2$; | $N_0 \times \widehat{N}_1/P_0 \times \widehat{N}_2/P_1 \xleftarrow{IFFT} \widehat{N}_0 \times \widehat{N}_1/P_0 \times \widehat{N}_2/P_1$; |
| $N_0/P_0 \times N_1 \times \widehat{N}_2/P_1 \xleftarrow{T} N_0/P_0 \times N_1/P_1 \times \widehat{N}_2$; | $N_0/P_0 \times \widehat{N}_1 \times \widehat{N}_2/P_1 \xleftarrow{T} N_0 \times \widehat{N}_1/P_0 \times \widehat{N}_2/P_1$; |
| $N_0/P_0 \times \widehat{N}_1 \times \widehat{N}_2/P_1 \xleftarrow{FFT} N_0/P_0 \times N_1 \times \widehat{N}_2/P_1$; | $N_0/P_0 \times N_1 \times \widehat{N}_2/P_1 \xleftarrow{IFFT} N_0/P_0 \times \widehat{N}_1 \times \widehat{N}_2/P_1$; |
| $N_0 \times \widehat{N}_1/P_0 \times \widehat{N}_2/P_1 \xleftarrow{T} N_0/P_0 \times \widehat{N}_1 \times \widehat{N}_2/P_1$; | $N_0/P_0 \times N_1/P_1 \times \widehat{N}_2 \xleftarrow{T} N_0/P_0 \times N_1 \times \widehat{N}_2/P_1$; |
| $\widehat{N}_0 \times \widehat{N}_1/P_0 \times \widehat{N}_2/P_1 \xleftarrow{FFT} N_0 \times \widehat{N}_1/P_0 \times \widehat{N}_2/P_1$; | $N_0/P_0 \times N_1/P_1 \times N_2 \xleftarrow{IFFT} N_0/P_0 \times N_1/P_1 \times \widehat{N}_2$; |

A very interesting set of papers proposes a different FFT algorithm that has lower global communication constants. It requires one all-to-all communications as opposed to three, and can be made up to 2× faster by introducing an approximation error in the numerical calculations. The algorithm was introduced in [38] and its parallel implementation discussed in [28]. Now it is part of the MKL library. It currently supports 1D FFT transforms only.

*Contributions.* We present AccFFT, a library that given an $N_0 \times N_1 \times N_2 \times \cdots$ matrix of values computes its Fourier Transform. The library supports the following features:

- hybrid MPI and CUDA parallelism,

- An overlapping all-to-all that reduces the PCIe overhead

- slab (1D) and pencil (2D) decomposition, and

- support for real-to-complex (R2C), complex-to-complex (C2C), and complex-to-real (C2R) transforms

- Single and double precision support

- Fast spectral operators

AccFFT uses CUFFT and FFTW for the FFT computations. AccFFT extends the single-node version of these two libraries to pencil decomposition for distributed memory FFTs. Both decompositions are necessary in order to ensure good performance across a range of MPI ranks. Slab decomposition limits the number of MPI ranks, $P$, to be less or equal to $N_0 = \max\{N_0, N_1, N_2\}$, and thus does not scale as well as the pencil decomposition [1]. To the best of our knowledge, AccFFT is the only open source code for distributed GPU transforms. The only other library that is currently maintained is FFTE, which has very limited features. It only supports complex-to-complex tranforms and requires commercial PGI compiler. However, our code is open source and does not require such compilers. Furthermore, AccFFT supports more transforms and our experimental comparisons show that our GPU code is faster than FFTE.

We present scaling results on Maverick (with K40 GPUs) at TACC, and on Titan at ORNL (with K20 GPUs). We use novel communication algorithms for GPUs to hide the overhead of moving data forth and back from the CPU. The library is open source and available for download [2] under GNU GPL version 2 license.

*Limitations.* There are several limitations in our library. We are not using non-blocking collective communications. The authors of [35] demonstrated that such an asyn-chronous approach can yield further speedups. Also Currently we do not support pruned FFTs or additional features such as Chebyshev approximations. Our implementation does not support inexact FFTs. Currently there is no support for hybrid floating point computation, but for larger FFTs it may be necessary.

---
[1]Typical values for $N_0$ range from 100s to 10,000s

**Algorithm 2:** Forward and backward FFT algorithms for general $d-1$ dimensional decomposition.

| **Input** : Data in spatial domain. <br> Layout: $N_0/P_0 \times \cdots N_{d-1}/P_{d-1} \times N_d$ <br> **Output**: Data in frequency domain. <br> Layout: $\widehat{N}_0 \times \cdots \widehat{N}_{d-1}/P_{d-2} \times \widehat{N}_d/P_{d-1}$ <br> $h = N_0/P_0 \times \cdots \times N_{d-1}/P_{d-1}$; <br> $h' = 1$; <br> **for** $i = d,...,1$ **do** <br> $\quad h \times \widehat{N}_i \times h' \xleftarrow{FFT} h \times N_i \times h'$; <br> $\quad H = h/(N_{i-1}/P_{i-1})$; <br> $\quad H' = h' * (\widehat{N}_i/P_{i-1})$; <br> $\quad H \times N_{i-1} \times H' \xleftarrow{T} h \times \widehat{N}_i \times h'$; <br> $\quad h = H; h' = H'$; <br> $h \times \widehat{N}_0 \times h' \xleftarrow{FFT} h \times N_0 \times h'$; | **Input** : Data in frequency domain. <br> Layout: $\widehat{N}_0 \times \cdots \widehat{N}_{d-1}/P_{d-2} \times \widehat{N}_d/P_{d-1}$ <br> **Output**: Data in spatial domain. <br> Layout: $N_0/P_0 \times \cdots N_{d-1}/P_{d-1} \times N_d$ <br> $h = 1$; <br> $h' = \widehat{N}_1/P_0 \times \cdots \times \widehat{N}_d/P_{d-1}$; <br> **for** $i = 0,...,d-1$ **do** <br> $\quad h \times N_i \times h' \xleftarrow{IFFT} h \times \widehat{N}_i \times h'$; <br> $\quad H = h * (N_i/P_i)$; <br> $\quad H' = h'/(\widehat{N}_{i+1}/P_i)$; <br> $\quad H \times \widehat{N}_{i+1} \times H' \xleftarrow{T} h \times \widehat{N}_i \times h'$; <br> $\quad h = H; h' = H'$; <br> $h \times N_d \times h' \xleftarrow{IFFT} h \times \widehat{N}_d \times h'$; |
|---|---|

**Algorithm 3:** Forward and backward FFT algorithm for slab decomposition.

| **Input** : Data in spatial domain. <br> Layout: $N_0/P \times N_1 \times N_2$ <br> **Output**: Data in frequency domain. <br> Layout: $\widehat{N}_0 \times \widehat{N}_1/P \times \widehat{N}_2$ <br> $N_0/P \times \widehat{N}_1 \times \widehat{N}_2 \xleftarrow{FFT} N_0/P \times N_1 \times N_2$; <br> $N_0 \times \widehat{N}_1/P \times \widehat{N}_2 \xleftarrow{T} N_0/P \times \widehat{N}_1 \times \widehat{N}_2$; <br> $\widehat{N}_0 \times \widehat{N}_1/P \times \widehat{N}_2 \xleftarrow{FFT} N_0 \times \widehat{N}_1/P \times \widehat{N}_2$; | **Input** : Data in frequency domain. <br> Layout: $\widehat{N}_0 \times \widehat{N}_1/P_0 \times \widehat{N}_2$ <br> **Output**: Data in spatial domain. <br> Layout: $N_0/P_0 \times N_1 \times N_2$ <br> $N_0 \times \widehat{N}_1/P_0 \times \widehat{N}_2 \xleftarrow{IFFT} \widehat{N}_0 \times \widehat{N}_1/P_0 \times \widehat{N}_2$; <br> $N_0/P_0 \times \widehat{N}_1 \times \widehat{N}_2 \xleftarrow{T} N_0 \times \widehat{N}_1/P_0 \times \widehat{N}_2$; <br> $N_0/P_0 \times N_1 \times N_2 \xleftarrow{IFFT} N_0/P_0 \times \widehat{N}_1 \times \widehat{N}_2$; |
|---|---|

*Outline of the paper.* In Section 2, we summarize our algorithms for CPU and GPU platforms, and discuss details of our optimizations. In Section 3, we present results of the numerical experiments.

## 2. Algorithm

In this section we discuss the AccFTT library's algorithms. First let us introduce some basic notation: $f$ is input array, $\widehat{f}$ is its Fourier transform, $P$ is the total number of MPI tasks, $N_0, N_1, N_2$ denotes the size of $f$ in x,y, and z direction, and $N = N_0 \times N_1 \times N_2$.

The discrete 3-D Fourier transform corresponds to a dense matrix-vector multiplication. However, the computational complexity can be reduced by using Cooley-Tukey algorithm to:

$$5N_0 N_1 N_2 (\log(N_0 N_1 N_2)).$$

The forward FFT maps space to frequency domain and the inverse FFT maps the frequency to space domain. The algorithms are the same up to a scaling factor, and have the same computational complexity[2].

For many scientific applications, $f$ does not fit into a single node and therefore the data needs to be distributed across several nodes. Two such distributions for a 3D array are the slab decomposition, in which the data is distributed in slabs and the pencil decomposition where each task gets a pencil of the data, as shown in Figure 1. To compute the FFT, each task has to compute its portion of FFTs, and then exchange data with other tasks. One can either use a binary exchange or a transpose (all-to-all) algorithm [16]. In this paper we focus on the latter.

---

[2]Note that typically FFT libraries do not apply scaling when forward FFT is computed and instead a full normalization is done when inverse FFT is performed. Therefore the complexity of the inverse would be slightly different than forward FFT, but the asymptotic behaviour would be the same.

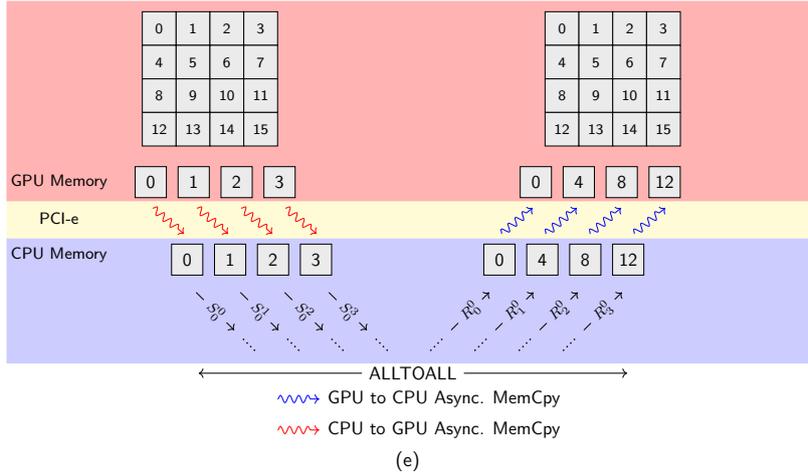

Figure 2: Demonstration of the GPU all-to-all. We interleave PCI-e transfers with send/recv operations to hide the overhead of sending data back/forth from the CPU. Here we are showing the alltoall process for the case of $P = 4$. Each element in the $4 \times 4$ matrix, represents a super element in the memory. $S_i^j$ denotes a send directive to process j from process i. Similarly, $R_i^j$ denotes receive message at process j sent from process i.

First we discuss the slab-decomposition, which is outlined in Algorithm 3. The input data is distributed in the first dimension over $P$ tasks, i.e. $N_0/P \times N_1 \times N_2$, which is referred to as a slab. Without loss of generality, let us assume that $N_0 = \max\{N_0, N_1, N_2\}$. In the limit of $P = N_0$, each task will just get a 2D slice of $f$ locally. If $P > N_0$ the slab decomposition cannot be used. In the forward algorithm, each task computes a $N_0/P$-batch of 2-D FFTs, each one having a size of $N_1 \times N_2$. Then an all-to-all exchange takes place to redistribute the data (indicated by the second step marked as $T$ in Algorithm 3. After this each task gets a slab of size $N_0 \times \widehat{N}_1/P \times \widehat{N}_2$, where the hats denote that the Fourier Transform has been computed across the last two dimensions. To complete the transform, each task can then compute a batch of $\widehat{N}_1/P \times \widehat{N}_2$ 1-D FFTs of length $N_0$. The inverse FFT can be computed in a similar fashion by reversing these steps. One advantage of our implementation is that the memory layout of the data in frequency space is the same as in the spatial one. This is different from, e.g., PFFT or FFTW's implementation, where the default (and faster) option, changes the memory layout from xyz to yxz. They provide an option which brings back the memory to the original xyz format, but at the cost of a local transpose [3].

Slab decomposition can be modified by decomposing the second dimension as well, which is known as pencil decomposition (Algorithm 1). In this approach each task gets a batch of 1-D pencils, local in the last dimension. That is the memory layout of the data in each task is $N_0/P_0 \times N_1/P_1 \times N_2$. The MPI tasks are mapped to a 2D matrix with $P_0$ rows and $P_1$ columns such that $P = P_0 \times P_1$. To compute the forward FFT, each task first computes a $N_0/P_0 \times N_1/P_1$ batch of 1-D FFTs of length $N_2$. This is followed by a block row-wise all-to-all communication step in which all the tasks in the same row call exchange to collect the second dimension of the array locally. In this step, one needs to redistribute a batch of $N_0/P_0$ matrices of size $N_1/P_1 \times N_2$. A naive implementation of this phase would lead to costly cache misses. As a result, we first perform a local packing of the data, so that all the batched noncontiguous data, are grouped together in a contiguous buffer. Then the all-to-all operation is performed, followed by an unpacking operation to get the data back into its correct format. Another approach is to use MPI data types to exchange non contiguous data. However, the latter would depend on how well the MPI compiler handles non-contiguous data. For consistency we do the packing and unpacking before the all-to-all calls.

---

[3]Note that this is different from global transposes. Here we are referring to the case where the global data layout is transposed. That is after passing the equivalent of TRANSPOSED_OUT flag for each library, by default they change the memory footprint of the data in frequency domain to yxz instead of xyz.

After the first all-to-all exchange, each task computes a batched 1-D FFT of size $N_1$ of its local data, which is now in the form of $N_0/P_0 \times N_1 \times \widehat{N}_2/P_1$. This is followed by an the all-to-all operation performed on a $N_0/P_0 \times N_1$ matrix with super-elements of size $\widehat{N}_2/P_1$ complex numbers. In this step, the data is indeed contiguous and no packing/unpacking is required. However, we do perform a local transpose, to change the memory layout from $N_1/P_0 \times N_0 \times N_2/P_1$ to $N_0 \times N_1/P_0 \times N_2/P_1$. This is done to have a consistent memory access pattern in both the spatial and frequency domain, which is desirable from a user's standpoint. To complete the forward FFT, a final batched 1-D FFT of length $N_0$ should be computed after this step. Now in the frequency space, each task owns the x pencil locally, while in spatial domain it owns the z dimension locally. This can be changed by performing two more all-to-all exchanges, but it is typically avoided as it can be done while the inverse FFT is being computed. The pencil decomposition algorithm can be extended to support n-dimensional tensors as shown in Algorithm 2.

It is well known that the most expensive part of distributed FFT is the communication phase [7], which adversely affects the scaling at large core counts. This has been verified in large scale runs on Stampede and Blue Waters [20]. This phase involves all-to-all exchanges, which is essentially transpose operation between a subgroup of tasks. As mentioned earlier, this exchange should be wrapped around a packing/unpacking phase to make the data contiguous in the memory. Generally the packing/unpacking phase accounts for less than 10% of the transpose time. This phase can be performed either by reshuffling of the data as done in P3DFFT, a local transpose as implemented in FFTW library, or eliminated by using MPI Data types [34]. However, the communication time dominates the cost of distributed FFT. The situation is even worse for the GPU, since the data has to be transferred forth and back to the CPU through PCIe, which is as expensive as the communication phase.

Recently, NVIDIA has introduced GPUDirect technology where GPUs on different nodes can communicate directly through the PCIe bus and avoid the CPU altogether. However, this feature requires special hardware support as well as a compatible OFED (OpenFabrics Enterprise Distribution). In the absence of GPUDirect, one option is to perform a blocking memcpy from GPU to CPU, use the transpose functions that are already implemented in the CPU code, and then copy back the results to the GPU. The packing and unpacking phases can still be performed on the GPU, as it can perform the reshuffling/local transposes much faster than on the CPU.

However, it is possible to hide the extra cost of memcpy by interleaving it into send and receive operations. Instead of copying all the data at once and then sending it, we divide the memcpy into chunks of the same size that each process has to send to other tasks. Each chunk is copied to a CPU buffer at a time, followed by an asynchronous send instruction. In this manner, the communication part of the CPU can start while the rest of the chunks are being copied. Since we post the asynchronous receive instructions beforehand, the receive operation can also happen with the device to host memcpy. Each received chunk can then be copied asynchronously back to the GPU Fig. 2.

For local FFT computations on GPUs, we use the CUFFT library from NVIDIA. One development was to implement a local transpose since the transpose on the NVIDIAs SDK libraries is not appropriate for the 3-D (or higher dimensioanl) FFTs since it doesn't support the correct stride and n_tuples. The second and the main contribution was to work around the limited bandwidth between the host CPU and GPU, and hide its overhead.

*Complexity Analysis.* The communication cost is $\mathcal{O}(\frac{N}{\sigma(p)})$, where $\sigma(p)$ is the bisection bandwidth of the network (for a hypercube it is $p/2$ [29]). The total execution time for an FFT of size $N$ on a hypercube can be approximated by:
$$T_{\text{FFT}} = \mathcal{O}\left(\frac{N \log N}{P}\right) + \mathcal{O}\left(\frac{N}{P}\right).$$
The first term represents the computation and the second the memory and communication costs. For a 3-D torus topology (such as the one used on Titan) the complexity becomes:
$$T_{\text{FFT}} = \mathcal{O}\left(\frac{N \log N}{P}\right) + \mathcal{O}\left(\frac{N}{P^{2/3}}\right).$$
For the GPU version this should also include the device-host communication costs. In [7] the authors give a detailed analysis in which cache effects and the local and remote memory bandwidth for GPUs and CPUs is taken into account. The basic point is that in strong scaling, the computation part becomes negligible and the overall wall-clock time will be dominated by the communication costs.

## 3. Numerical experiments

In this section we report the performance of AccFFT and give details regarding our implementation and the different problem sizes used for evaluating the library.

*Computing Platforms*

The tests are performed on the following platforms:

- The **Maverick** system at TACC is a Linux cluster with 132 compute nodes, each with dual 10-core 2.8GHz Intel Xeon E5 (Ivy Bridge) processors with 13GB/core of memory equipped with FDR Mellanox InfiniBand network. Each of its 132 nodes is equipped with a K40 GPU.

- The **Titan** system is a Cray XK7 supercomputer at ORNL. Titan has a total of 18,688 nodes consisting of a single 16-core AMD Opteron 6200 series processor, for a total of 299,008 cores. Each node has 32GB of memory. It is also equipped with a Gemini interconnect. In addition, all of Titan's 18,688 compute nodes contain an NVIDIA Tesla K20 GPU/

- The **Stampede** system at TACC is a Linux cluster consisting of 6400 compute nodes, each with dual, eight-core processors for a total of 102,400 available CPU-cores. Each node has two eight-core 2.7GHz Intel Xeon E5 (Sandy Bridge) processors with 2GB/core of memory and a three-level cache. Stampede has a 56Gb/s FDR Mellanox InfiniBand network connected in a fat tree configuration.

*Implementation Details.* All algorithms described in this work were implemented using C++, OpenMP and MPI. The only external libraries used where the MPI, FFTW, and CUFFT. On Titan, we used the GCC compiler and the CRAY-MPICH libraries. On Stampede and Maverick we used the Intel compilers and the Intel MPI library We compare our GPU code with FFTE library (version 6.0). The GPU code for FFTE library is written in Fortran and requires the commercial PGI compiler, and cuFFT for its FFT computations on the GPU. All the libraries were compiled with the MEASURE planner flag where applicable. This flag is used to tune the libraries to the machine used. All results were computed in double precision and with pencil decomposition.

*Parameters in the Experiments* The parameters in our runs are the problem size $N_0, N_1, N_2$ and the number of tasks $P$. In most of the tests, we use $N_0 = N_1 = N_2$. The exception are two tests in which we test the library

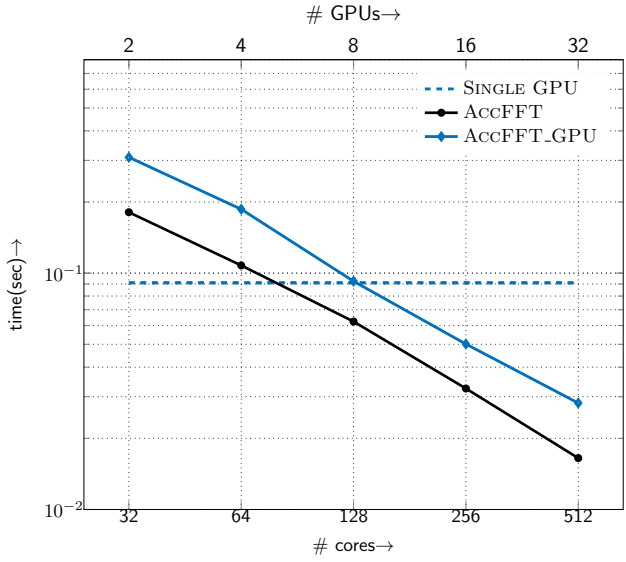

(a) R2C strong scaling for $N = 256 \times 512 \times 1024$

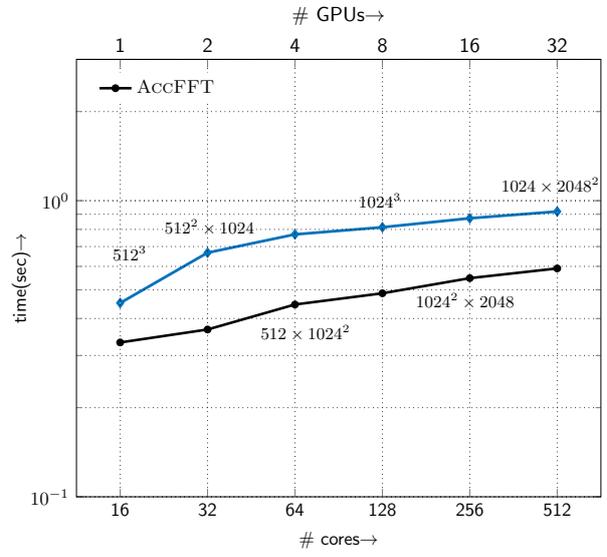

(b) R2C weak scaling

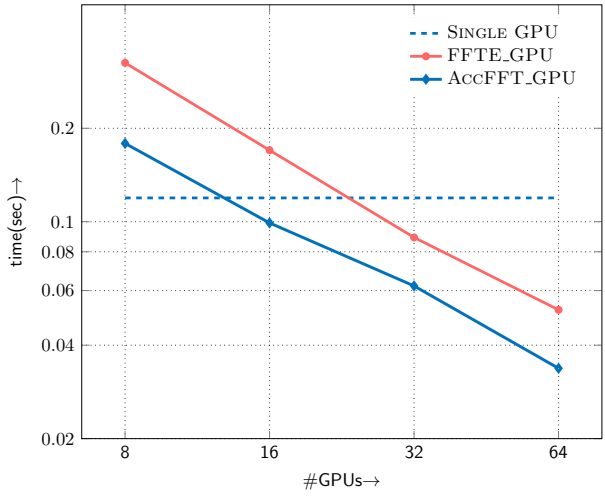

(c) C2C strong scaling for $N = 256 \times 512 \times 1024$

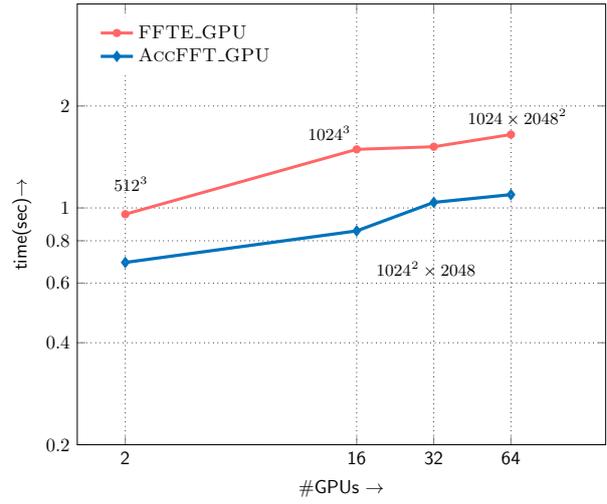

(d) C2C weak scaling

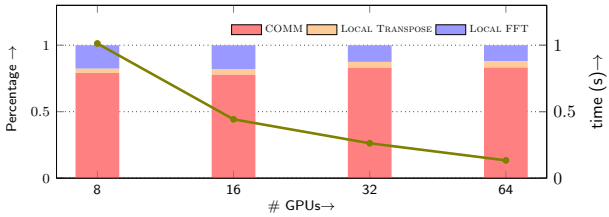

(e) R2C time break down for $N = 1024^3$

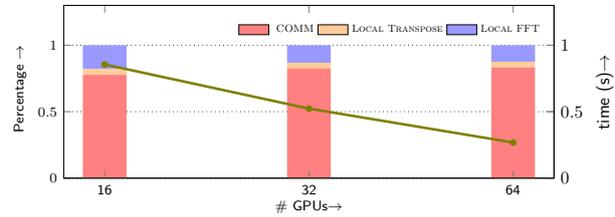

(f) C2C time break down for $N = 1024^3$

Figure 3: *Scaling results of AccFFT performed on Maverick. (a) strong scaling result for a non-structured R2C transform of size $N = 256 \times 512 \times 1024$, (b) R2C weak scaling, (c) C2C Strong scaling for $N = 256 \times 512 \times 1024$ and comparison with FFTE's GPU code, (d) C2C weak scaling, (e) breakdown of timings for R2C transform of size $N = 1024^3$, (f) breakdown of timings for C2C transform of size $N = 1024^3$.*

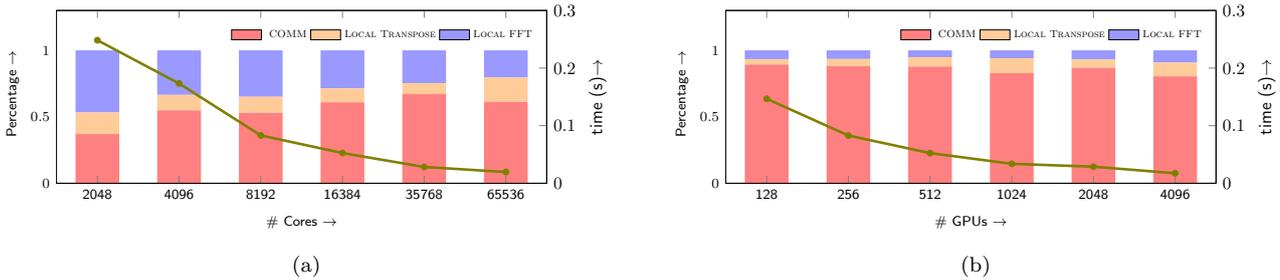

Figure 4: *Strong scaling results for AccFFT performed on Titan. Timings in seconds are given for R2C transform of size $N = 1024^3$ on CPU and GPU (left/right). The breakdown of the total time in terms of local FFT computations, packing and unpacking (local transpose), and the communication times is given for different core counts.*

with non well-structured matrices, and a 4D test case. Except otherwise indicated, we use 16 MPI tasks per node for CPU runs and 2 MPI tasks per node for GPU tests. We use R2C to denote real-to-complex FFTs and C2C to denote complex to complex. Roughly speaking, the C2C transform has double the computation and communication compared to R2C transform. All timings are in seconds.

*Experiments* First we examine the performance of our code on the TACC systems, and then we discuss the results on Titan.

• In the first experiment we present scaling tests on Maverick. The strong scaling of the CPU and GPU code for a size of $N = 256 \times 512 \times 1024$ is shown in (Fig. 3 (a)). The GPU code scales similarly to the CPU code, however it is about $2\times$ slower. One reason for this is that we are comparing 16 CPU cores vs 1 GPU at each node. Local FFT computations scale almost perfectly as the cores are increased, so the advantage that the GPU has for its fast FFT computation would become negligible. The second point, is that part of the CPU communication occurs inside the same node, which is much faster than two GPUs communicating from different nodes. Weak scaling analysis shows the same trend, as shown in Fig. 3 (b).

There is currently no open source GPU code that supports pencil decomposition, other than digpufft and FFTE. The digpufft library, which is built on top of P3DFFT, is no longer maintained and our attempts to run it were not successful. However, FFTE library is maintained and supports C2C transforms with pencil decomposition [36]. Both our library as well as FFTE use cuFFT for local FFT computations. However, FFTE relies on the commercial PGI complier for the communication phase and does not do any optimizations. The two libraries are compared in Fig. 3 (c-d). AccFFT is consistently faster in both the strong and weak scaling tests. Moreover, AccFFT supports other transforms and is not just limited to C2C.

The break down of the timings for R2C and C2C transforms for $N = 1024^3$ is shown in figures 3 (e-f). Again the communication phase dominates the cost.

• Now we switch to Titan, where we consider the strong scaling of GPU versions of the code. We also present CPU results on Titan with 2 MPI tasks per node (1 tasks per NUMA). The goal is to compare the codes where the communication pattern between the CPU code and the GPU code is similar (that is there is no intra node acceleration that the GPU does not have). This allows us to see how effective does the PCIe overlapping works. The results are showin in Fig. 4 for up to 4096 GPUs and 65K cores. The efficiency for the largest CPU run is 40% for a $32\times$ increase in the core count (2,048 to 65K cores). The GPU code achieves 26% efficiency for a $32\times$ increase in the number of GPUs. The GPU code compared to 2 CPU cores is obviously faster as expected. An interesting observation is that the communication times are comparable which is becaues of the overlapping of PCIe excahnges. Although we did not have a chance to test 16 MPI tasks per node for the CPU case on Titan, but the CPU time is expected to become faster compared to the GPU code similar to the Maverick results.

• Finally, as a proof of concept we show how the code can be used for high dimensional transforms. In certain applications such as in signal processing, one needs to compute FFT of a 4D array, where the last dimension corresponds to time. Figure ?? shows the strong scaling of the CPU code for a C2C transform of size $N = 512 \times 256 \times 128 \times 64$. Nothing changes in our communication algorithms for high dimensional transforms for either the CPU or the GPU code. [4]

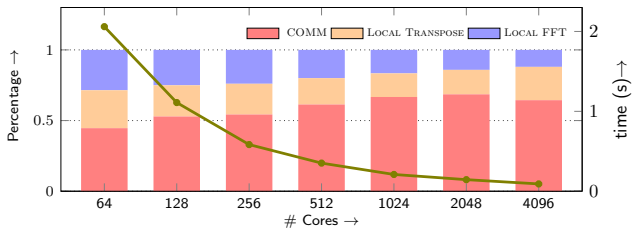

Figure 5: *CPU strong scaling results of AccFFT using the pencil decomposition algorithm for a four dimensional problem size of $512 \times 256 \times 128 \times 64$ on Stampede. Wall-clock time is given for different core counts for a forward C2C FFT.*

### 4. Conclusions

We presented AccFFT, a library for distributed memory FFTs with several unique features: distributed GPU calculations, communication overlap using pipelining for the GPU version, and support for real and complex transforms.

The performance of the library was tested on three different machines: The Stampede and Maverick systems at TACC, as well as the Titan system at ORNL. The largest test corresponds to 65K cores as well as 4,096 GPUs of Titan. To the best of our knowledge, the latter is the only open source library supporting pencil decomposition for GPUs. One of the observations of this work is that parallel GPU is not faster when compared to parallel CPU, in cases where there are multiple CPU cores per node. The main reason for this is that local FFT computations scale almost perfectly. So using more MPI tasks per node would significantly reduce the local FFT time for CPU. Moreover, the communication phase of the CPU would benefit from fast intra-node exchanges through shared memory. This is not the case for machines which habe one GPU per node. In that case two GPUs have to communicate through the network which is much costlier.

This work is by no means complete with these results. Using non-blocking collectives can further accelerate the calculations, if the user interleave other computations while the FFT communication is completing. Other possibilities include distributing the data to both the CPU and GPU on each node. This has been shown to be an effective strategy on single node computations [40]. Another possibility is to extend the method to Xeon Phi accelerators.

---

[4]This feature has not been added to the public repo the library yet.